\documentclass{elsart}
\usepackage{amsmath,epsfig,amsfonts,euscript,amssymb,latexsym}

\begin{document}

[{\scriptsize A.E.\ Motter, M.A.\  Mat\'{i}as, J.\ Kurths, and E.\ Ott,
Physica D {\bf 224} (2006), pp. vii-viii}]

\begin{frontmatter}

\title{Dynamics on Complex Networks and Applications}

\author[A]{Adilson E. Motter\thanksref{corres_auth}},
\author[M]{Manuel A. Mat\'{i}as},
\author[J]{J\"urgen Kurths}, and
\author[E]{Edward Ott}

\address[A]{Department of Physics and Astronomy, Northwestern University, Evanston, IL 60208, USA}
\address[M]{Instituto Mediterr\'aneo de Estudios Avanzados, IMEDEA (CSIC-UIB), E-07122 Palma de Mallorca, Spain}
\address[J]{Institute of Physics, University of Potsdam PF 601553, 14415 Potsdam, Germany}
\address[E]{Department of Physics and Department of Electrical and Computer Engineering, University of Maryland,
            College Park, Maryland 20742, USA}

\thanks[corres_auth]{Corresponding author: {\tt motter@northwestern.edu}, +1-847-491-4611 (TEL), +1-847-491-9982 (FAX)}

\date{November 1, 2006}

\begin{abstract}
At the eight-year anniversary of Watts \& Strogatz's work on the collective
dynamics of small-world networks and seven years after Barab\'asi \& Albert's 
discovery of scale-free networks, the area of dynamical processes on complex
networks is at the forefront of the current research on nonlinear dynamics and
complex systems.
This volume brings together a selection of original contributions in complementary
topics of statistical physics, nonlinear dynamics and biological sciences, and 
is expected to provide the reader with a comprehensive up-to-date representation
of this rapidly developing 
area.
\end{abstract}

\begin{keyword}
  complex systems \sep nonlinear dynamics \sep statistical physics
  \PACS 05.10.-a \sep 05.45.Xt \sep 89.75.-k \sep 87.18.Sn
\end{keyword}

\end{frontmatter}

About collective behavior, Philip Anderson already said in 1972 ``more is different" 
\cite{anderson:1972}.
But how different? 
This question is gaining new momentum with the emergence of fields of research that are
treating
in great detail the properties of individual components of complex systems, such 
as genes and proteins in a cell or neurons in the brain. The increasing availability of
information about 
the components of systems calls
for a parallel development of a system-level
analysis capable of describing the integrated collective behavior. 

The theory of complex networks seems to offer an appropriate framework for such a large-scale 
analysis in a representative class of complex systems, with examples ranging from cell biology
and epidemiology to the Internet 
\cite{watts:1999,amaral:2000,strogatz:2001,ab:2002,dm:2002,bs:2002,newman:2003}.
The research in this area has been fueled by the discovery of universal structural properties
in real-world networks and the theoretical understanding of evolutionary laws governing the emergence
of these properties 
\cite{ws:1998,ba:1999}. 
However, implicit in the 
study of the function of such networks
is the idea that 
they sustain
dynamical processes. 
Therefore, along with the study of purely structural
and evolutionary properties, there has been an increasing interest in the interplay
between the dynamics and the structure of complex networks 
\cite{strogatz:2001,pv:2004,b:2006,nbw:2006}.

This interest is well motivated since most biological, social and technological complex systems
are inherently dynamic. In these contexts, time-dependent phenomena are intimately related to the
performance of the system, as exemplified by traffic congestion, cascading failures, and synchronization
of biological oscillators. 

The idea that dynamical processes could be strongly influenced by the structure of an underlying network was
already suggested by Watts and Strogatz in their work on small-world networks \cite{ws:1998} using epidemic
spreading as an example process. In the context of an epidemic spreading, the importance of the network structure
became even more evident after Barab\'asi and Albert's work on scale-free networks \cite{ba:1999} and subsequent
study by Pastor-Satorras and Vespignani on the absence of epidemic threshold in such networks  \cite{sv:2001}.
The role of the network structure is further emphasized by the presence of communities  \cite{gn:2002}, correlations  
\cite{newman:2002}, patterns of weighted connections \cite{bbsv:2004}, and other nontrivial structures in many real-world networks that had not been 
anticipated from the classical random graph theory of Erd\H{o}s and R\'enyi \cite{er:1960}.
These alone could serve as a good motivation for the study of dynamics on complex networks.
However, there is more to it.

A salient property of most dynamical processes in complex systems is their almost unavoidable nonlinearity. 
Part of the recent interest in the study of dynamics on complex networks comes from the understanding that
techniques and expertise developed in the study of nonlinear dynamics and chaos can be useful in the study of
such nonlinear systems. A prime example of this is the master stability framework previously introduced by
Pecora and Carroll \cite{pc:1998} to study synchronization of coupled chaotic oscillators. The same framework
has been subsequently applied to separate the contribution of the network structure
encapsulated
in the eigenvalues of the coupling matrix in the study of network synchronization \cite{bp:2002,nmlh:2003}.
Similar eigenvalues also govern the influence of the network on a number of other dynamical phenomena,
including diffusion \cite{mzk:2005} and the emergence of coherent behavior in general \cite{roh:2006}. 
These approaches thus help determine not only how the system depends on its parts, but also how it depends on
the {\it way} the parts are linked together, and this offers a solid glimpse into how different ``more can be".

Because of all this, there is now a fast developing science of dynamics on complex networks that has as an
essential underpinning a strong interaction among different disciplines and between theory and applications. 
To help crystallize this emerging field as a promising area of interdisciplinary research, we have coordinated
the four-week International Seminar and Workshop ``Dynamics on Complex Networks and Applications" at the
Max Planck Institute for the Physics of Complex Systems, Dresden, February 06 - March 03, 2006, which was attended
by a hundred physicists, mathematicians,
and life scientists, among others.

In this Focus Issue, we bring together a selection of original contributions from leading experts 
in nonlinear dynamics, statistical physics, and biological sciences. This combination of contributions
from experts primarily working in different fields is a unique and outstanding aspect of this Focus Issue,
which is expected to enhance communication across disciplines, consolidate a common scientific language,
and foster the fast growing area of dynamics on complex networks. 

The volume is organized as follows. The first section consists of contributions on structural properties
of complex networks, and includes new results on clique counting, $k$-core percolation, and statistical
significance of community detection. The second section comprises a number of contributions on dynamics
on complex networks, with emphasis on synchronization and coherent behavior. These works present new 
results on the optimization of dynamical performance and structural cost, relations between functional
and structural communities, synchronization in networks of oscillators, spatio-temporal dynamics, and 
network transport. The third section includes contributions on social and human-generated networks where
the evolution of the network is coupled to an agent-driven dynamical process. The last section includes 
contributions on applications to gene regulatory networks, genetic similarity, and neuronal networks.

{\bf Acknowledgment:}
We would like to thank the MPIPKS for sponsoring the conference that culminated with this publication,
Professor Doelman and other members of {\em Physica D} editorial board for assisting us in preparing this volume,
and each of the authors for their truly valuable contributions.

\end{document}